\documentclass[a4paper,10pt]{article}
\usepackage{epsfig,cite,subfigure}
\begin{document}
%
\title{\textbf{
Alignment of nematic liquid crystals on mixed Langmuir-Blodgett 
mono-layers
}}
\author{
V. S. U. Fazio, L. Komitov and S. T. Lagerwall\\
\textit{\small{Department of Microelectronics \& Nanoscience,}}\\
\textit{\small{Chalmers University of Technology
and G\"oteborg University,}}\\
\textit{\small{S-41296 G\"oteborg, Sweden}}
}
\date{}
\maketitle
\begin{center}
\begin{minipage}{\textwidth}
\begin{abstract}
\noindent
Mono-layers of stearic and behenic acids
and mixtures of them in different
proportions, deposited with the Langmuir-Blodgett (\textsf{LB}) technique,
were used to study the alignment and the alignment dynamics
in nematic liquid crystal cells.
A relaxation process from a splay-bend flow induced 
metastable orientation to the homeotropic one occurs.
The lifetime of the metastable state was found to depend on the 
mono-layer composition.
The transition between the homeotropic and the
conical anchoring was found to be irreversible in the case of the
mixed aligning mono-layers: on cooling from the isotropic phase
a quasi-planar nematic state (schlieren texture) appears.
It is stable in a range of a few degrees below the clearing point
and, on decreasing the temperature, relaxes to the homeotropic state 
in form of expanding domains.
\end{abstract}
\bigskip
\end{minipage}
\end{center}
%
%
%
%


\section*{Introduction}
The Langmuir-Blodgett (\textsf{LB}) technique enables the
deposition of organic aligning films with controlled molecular
order and thickness similar to but much thinner that those often used as
aligning films in liquid crystals cells. 
They allow a very good homeotropic alignment over large areas.

\begin{figure}
\begin{center}
\epsfig{file=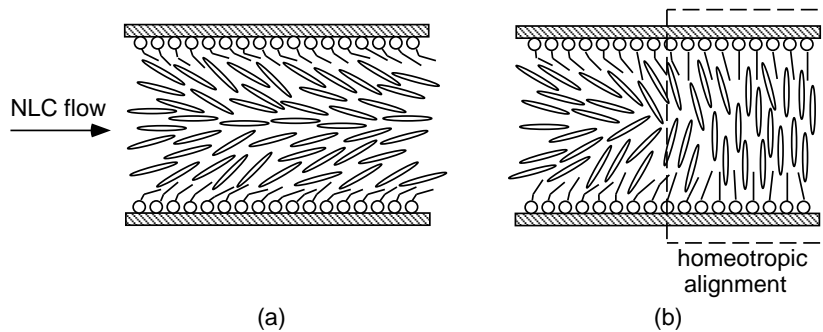,width=0.9\textwidth}
\caption{
\small{
(a) During filling the \textsf{LB} mono-layer seems to be strongly 
influenced by the flow with the chains distorted in the filling
direction.
A splay-bend deformation is induced in the \textsf{NLC} by
the flow.
(b) When the flow stops the \textsf{LB} film and the splay-bend
deformed liquid crystal both contribute to the relaxation towards the
homeotropic state{\protect\cite{FazKomLag97a}}.
\label{cellfilling_old}
}}
\end{center}
\end{figure}

Recently\cite{FazKomLag97a} we have shown that for pure
stearic (\textsf{C18}) and behenic (\textsf{C22}) aligning mono-layers
during the process of filling of the cell the nematic liquid crystal
(\textsf{NLC}) molecules are preferably aligned along the filling 
direction: the molecules in the centre of the cell gap are essentially 
parallel to the substrate and a splay-bend deformation in the 
\textsf{NLC} molecules is induced by the presence of the aligning layer
(see Figure \ref{cellfilling_old}). 
As soon as the flow stops on complete filling, 
domains of homeotropic alignment nucleate at the
edges of the cell and continuously grow until the whole sample 
becomes homeotropic.
The relaxation of the splay-bend deformation 
into the homeotropic state is partially a result of the elastic
relaxation of the splay-bend deformation, but seems to have a contribution 
also from the 
\textsf{LB} film itself which is distorted by the flow
and relaxes to the equilibrium state, where the mono-layer molecules
are in the upright position.
Since the lifetime of the splay-bend state was found to be
very similar for \textsf{C18} and \textsf{C22}, depending only
on the cell thickness, the r\^ole of the \textsf{LB} mono-layers 
in the process is much less pronounced than that of the
liquid crystal.

In this work mixed \textsf{C18}/\textsf{C22} mono-layers were used 
for investigating the alignment dynamics in nematic liquid crystal cells.
The alignment was studied during the filling process and pursued during the
relaxation to the equilibrium state.
The temperature stability of this state was also investigated.
\section*{Experiment}
1\,mM solutions of stearic, behenic, or stearic/behenic mixtures
in Merk chloroform were spread onto the surface of ultrapure
Milli-Q water in a \textsf{LB} trough.
Then the mono-layers were compressed at a rate of
0.3$\times$10$^{-3}$\,(nm$^{2}$s$^{-1}$molecule$^{-1}$) until the 
desired pressure was reached.

We used indium tin oxide (\textsf{ITO}) coated glasses as substrates.
They were first cleaned\cite{FazKomLag97a} and then immersed in 
the sub-phase before spreading the mono-layer.
The transfer onto the glass occurred during the extraction of the
glass from the sub-phase (10\,mm/min).

The glass plates were cut and assembled in sandwitch cells spaced with 
15\,$\mu$m polyester films supplied by Milar.
Care was taken do not touch, and thus contaminate, the inside
surface of the cells.

The cells were capillary filled with MBBA (Aldrich) at room
temperature. 
The observations on them were made with a polarizing microscope
in orthoscopic as well as conoscopic regime with the sample insered
in a hot stage between crossed polarizers.
The microscope was also equipped with a video-camera connected to
a computer.

\section*{Results}

\subsection*{Isotherms}
\begin{figure}
\begin{center}
\epsfig{file=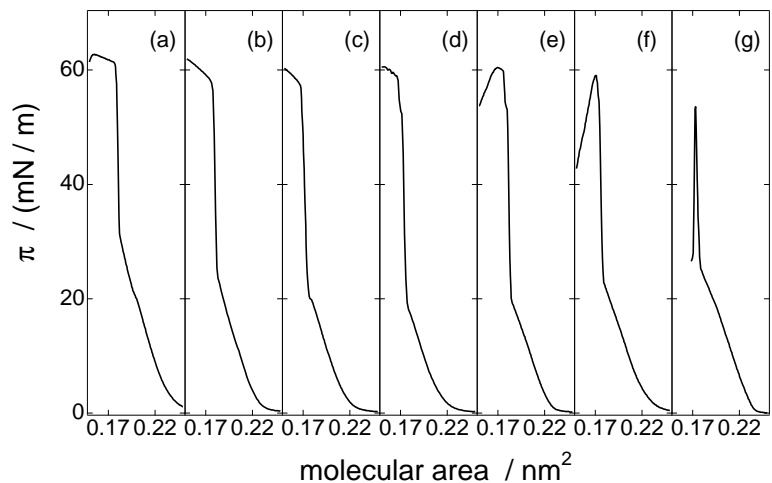,width=0.9\columnwidth}
\caption{\small{
\label{isoterme}
Surface pressure versus area per molecule isotherms for
stearic/behenic acid mixtures.
(a) 0:100, 
(b) 20:80,
(c) 30:70,
(d) 50:50,
(e) 60:40,
(f) 70:30,
(g) 100:0.
}}
\end{center}
\end{figure}

In Figure \ref{isoterme} the surface pressure versus
molecular area isotherms for films of pure stearic 
acid (\textsf{C18}), pure behenic acid (\textsf{C22}), 
and a number of \textsf{C18}/\textsf{C22}
mixtures with different \textsf{C18}:\textsf{C22}
ratios are shown.
%

To quantitatively study the molecular arrengement of the mixed
\textsf{C18}/\textsf{C22} mono-layers we followed Gaines
\cite{Gaines}.
Mono-layers in which the components are immiscible may be thought as made
up of two separate mono-layers in equilibrium with each other.
The area of the film will be the sum of the areas of the
separate films, or
\begin{equation}
A_{12} = N_{1} A_{1} + N_{2} A_{2},
\label{uno}
\end{equation}
where $A_{12}$ is the mean molecular area of the
two-component film, $A_{1}$ and $A_{2}$ are the molecular areas of
the two single component films at the same surface pressure, and 
$N_{1}$ and $N_{2}$ are the mole fractions of the two components.
Any deviation from (\ref{uno}),
also called \lq\lq line of no interaction\rq\rq,
provides evidence of miscibility 
as well as some sort of molecular interaction in the film
\protect{\cite{Gaines}}.

\begin{figure}
\subfigure[
\label{meanmolarea}
]
{\epsfig{file=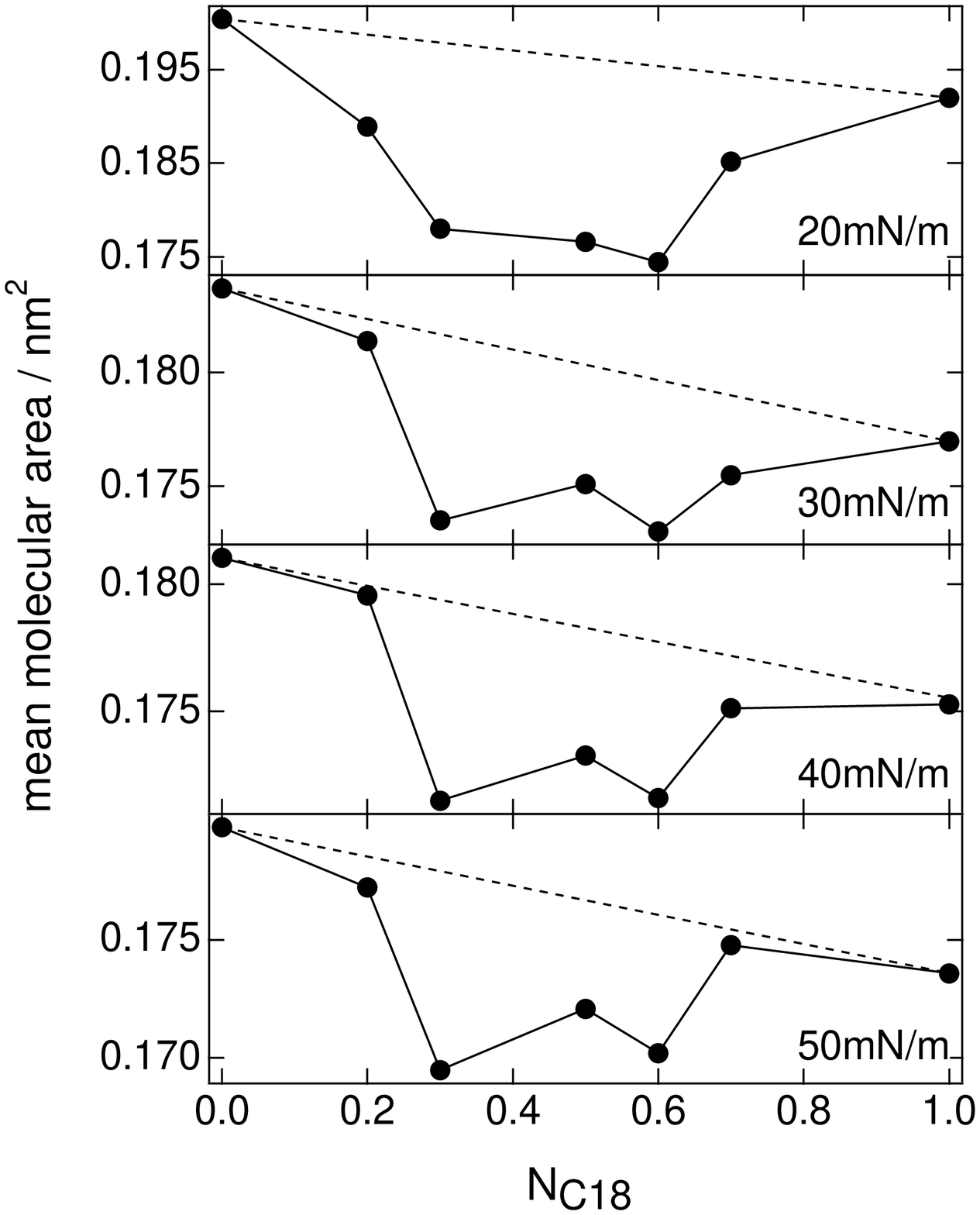,width=0.48\textwidth}}
\hspace{0.03\textwidth}
\subfigure[
\label{excessfre}
]
{\epsfig{file=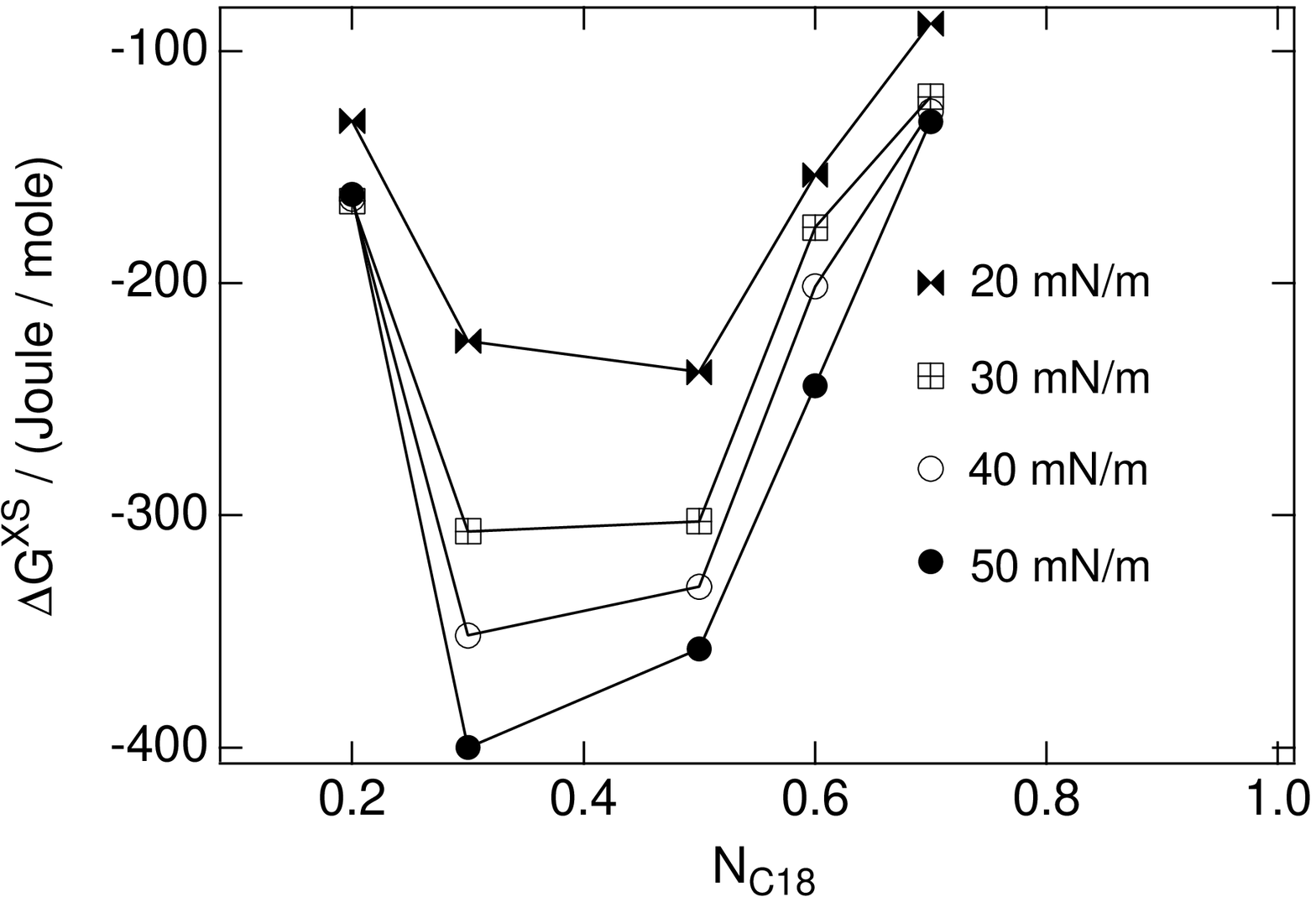, width=0.48\textwidth}}
\caption{
\small{
(a) Average molecular area of the \textsf{LB} film as a function of the
molar fraction of \textsf{C18}. 
The dashed line is the \lq\lq line of no interaction\rq\rq\, and
the dots are the experimental values.
The fact that all the experimental points lie below the 
\lq\lq line of no interaction\rq\rq\, indicates that the two
compounds 
mix{\protect\cite{Durham58,Albrecht-misc}}.
(b) Excess free energy of mixing calculated from the
isotherms in Figure \ref{isoterme} as a
function of the molar fraction of \textsf{C18} for four different
values of the $\pi^{*}$ pressure (cfr. eqn.({\protect\ref{due}})).
The values are all negative indicating 
that 
\textsf{C18} and \textsf{C22} 
mix.
}}
\end{figure}

In Figure \ref{meanmolarea} the mean molecular area in the mixed 
film is plotted versus the mole fraction of \textsf{C18}
for four different surface pressures.
In the figure 
the dashed line represents the \lq \lq line of
no interaction \rq\rq, and the dots
are the experimental values.
The fact that all the experimental points lie below the line 
of not interaction indicates that the two
components 
mix\cite{Durham58,Albrecht-misc}.

Once we established that the components of the mono-layer 
have a tendency to mix 
we may also calculate the excess free energy of mixing for a given 
surface pressure $\pi^{*}$\cite{Gaines},
\begin{equation}
\Delta G_{\mbox{\tiny{XS}}} =
\int_{0}^{\pi^*} (A_{12} - N_{1} A_{1} - N_{2} A_{2}) d \pi,
\label{due}
\end{equation}
which can be done directly from the $\pi-A$ curves of the
pure and the mixed mono-layers.
For a truly mixed mono-layer $\Delta G_{\mbox{\tiny{XS}}}$ must be negative.
In Figure \ref{excessfre} $\Delta G_{\mbox{\tiny{XS}}}$ 
is plotted as a function
of the \textsf{C18} fraction for four different surface 
pressures.       
The values of the excess free energy of mixing are all negative, 
confirming the already observed tendency of the two components to mix.

\subsection*{Alignment}
The cells were capillary filled with \textsf{MBBA} at room temperature
(where \textsf{MBBA} is in the nematic phase).
As we have shown\cite{FazKomLag97a},
during filling the chains of the molecules constituting the
\textsf{LB} film are distorted by
the flow and a splay-bend deformation is induced in the \textsf{NLC}.
As soon as the flow stops the 
distorted \textsf{LB} film and the splay-bend
deformed liquid crystal both contribute to the relaxation towards the
homeotropic state.

\begin{figure}
\subfigure[]{\epsfig{file=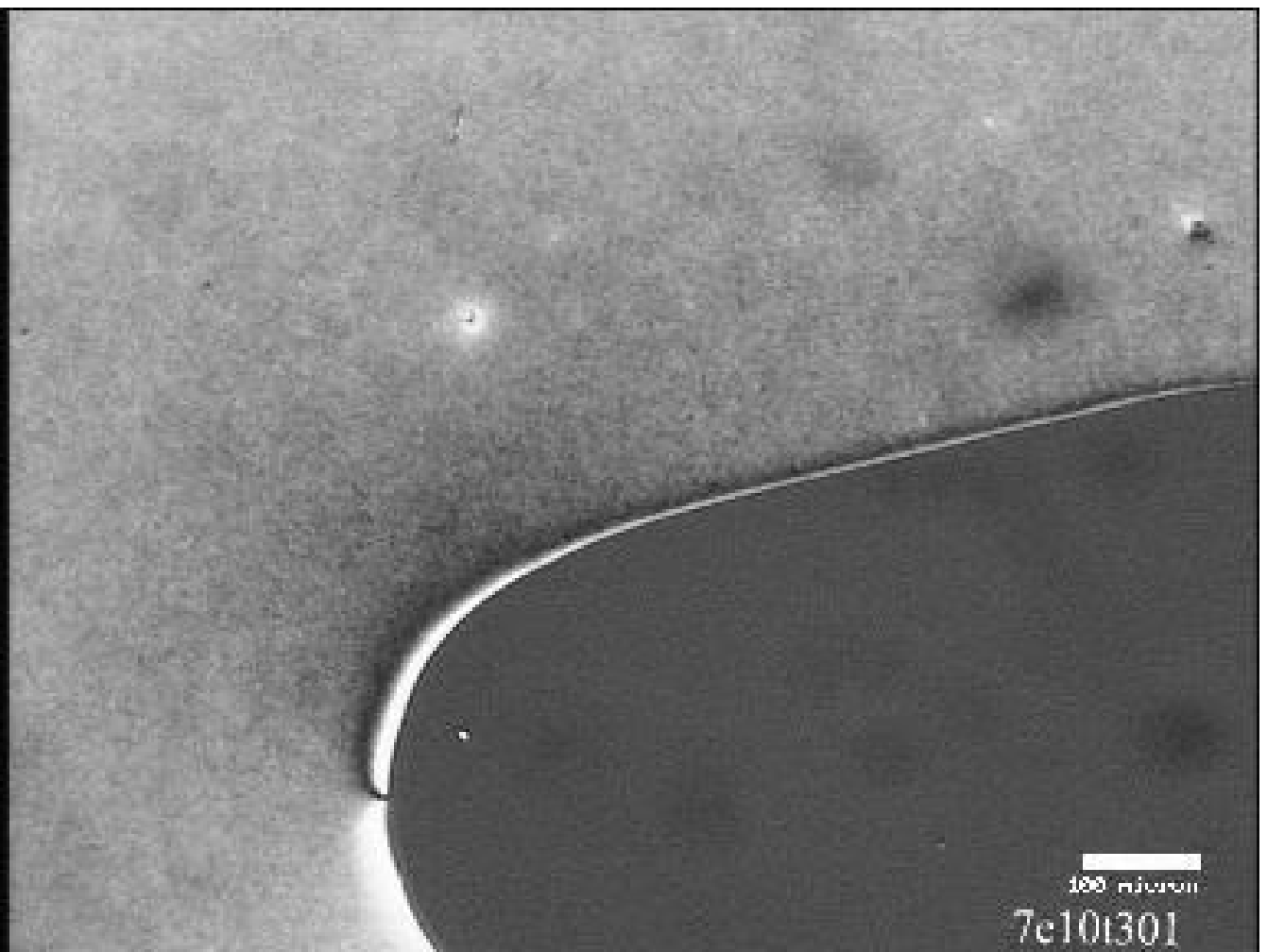,width=0.49\textwidth}}
\subfigure[]{\epsfig{file=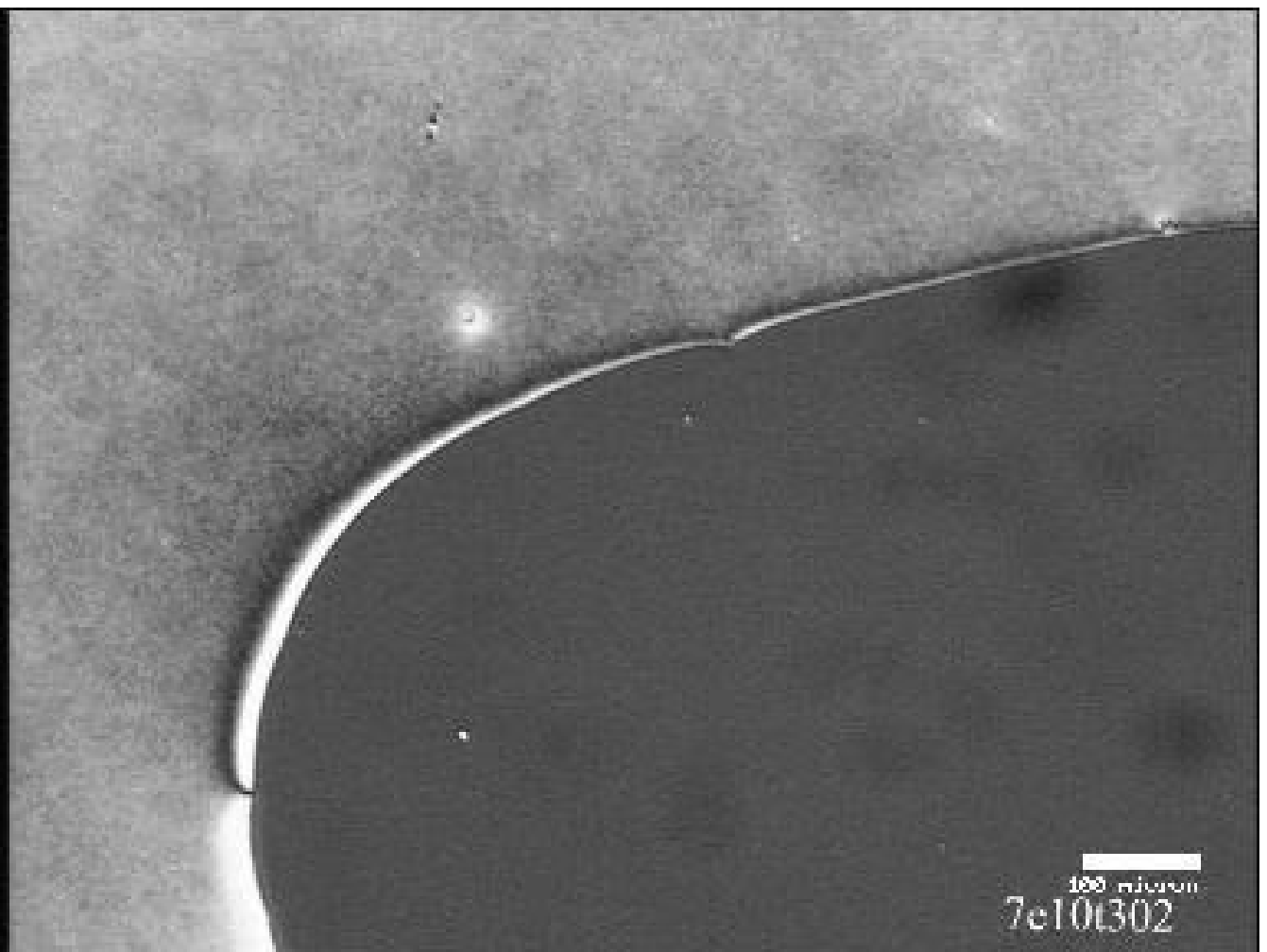,width=0.49\textwidth}}
\protect\caption{
\small{
Cell between crossed polarisers.
The \textsf{LB} aligning film is a \textsf{C18}:\textsf{C22}
50:50 mono-layer; the thickness of the cell is 15.2\,$\mu$m.
The cell is completely filled with \textsf{MBBA}, the
\textsf{NLC} flow has ceased, and the homeotropic domains (dark)
expand in the quasi-planar domain (light).
The two pictures were taken with a time interval of 20\,s.
\label{expand}\sloppy
}}
\end{figure}

An example of how the homeotropic domains expand in the cell is given
in Figure \ref{expand}.

\begin{figure}
\begin{center}
\epsfig{file=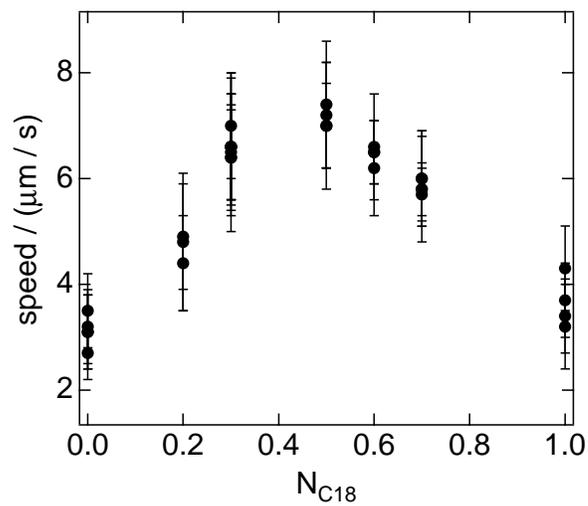,width=0.7\textwidth}
\caption{
\small{
Speed of expansion of the homeotropic domains as a function
of the \textsf{C18} fraction in the aligning mono-layer.
The cell thickness is about 15\,$\mu$m.
The homeotropic domains expand faster for the mixed aligning
mono-layers than for the pure ones.
The composition \textsf{C18}:\textsf{C22} 50:50 gives
the highest speed which decreases on increasing the quantity of one
or the other fatty acid.
\label{speed}
}}
\end{center}
\end{figure}

We measured the speed with which the homeotropic domains expand in the
cells taking pictures at fixed time intervals
for the different aligning layers.
The speeds were calculated as the area covered by 
the front of a homeotropic
domain in a certain time interval, divided by the length of the front and
the time interval.
The results are shown in Figure \ref{speed}.
The fact that the lifetime of the splay-bend state depends on the mono-layer
composition indicates that that the relaxation process is not only due
to the deformed liquid crystal, but also to the \textsf{LB} film.

\begin{figure}
\begin{minipage}{\textwidth}
\parbox[b]{0.5\textwidth}{
\epsfig{file=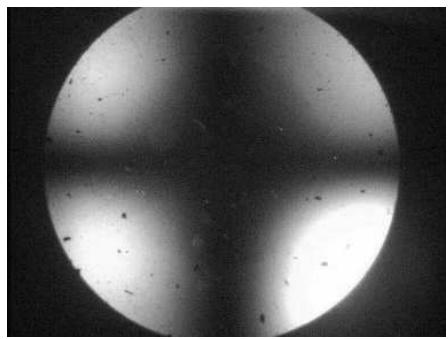, width=0.48\textwidth}
} 
\parbox[b]{0.49\textwidth}{
\protect\caption{\small{
Conoscopic figure of a cell with \textsf{C18}:\textsf{C22} 50:50 
aligning layer.
The cell is 15\,$\mu$m thick.
The homeotropic alignment is very good.
}}
{\label{conosco50}\sloppy }
}
\end{minipage}
\end{figure}

The quality of the homeotropic alignment was studied by conoscopy.
For a uniaxial compound ordelly oriented with the c-axis
perpendicular to the glass plates, the conoscopic figure
should consist of a black cross in a white background\cite{Ehlers}.
In Figure \ref{conosco50} the conoscopic figure of a
sample with  \textsf{C18}:\textsf{C22} 50:50 aligning layer 
is shown.
In Table \ref{alignment} the results of the conoscopy for all
the different aligning layers are summarised.
\begin{table}
\caption{
\small{
Alignment obtained for the different
compositions of the \textsf{LB} mono-layers.
}
\label{alignment}
}
\begin{center}
\begin{tabular}{cc}
\vspace{0.5mm}\\
\hline
\hline
\textsf{LB} film composition & Alignment \\
\hline
\textsf{C18}:\textsf{C22} \,\, 0:100 & Not well defined \\
\textsf{C18}:\textsf{C22} \,\, 20:80 & Not very good homeotropic \\
\textsf{C18}:\textsf{C22} \,\, 30:80 & Good homeotropic \\
\textsf{C18}:\textsf{C22} \,\, 50:50 & Very good homeotropic \\
\textsf{C18}:\textsf{C22} \,\, 60:40 & Very good homeotropic \\
\textsf{C18}:\textsf{C22} \,\, 70:30 & Good homeotropic \\
\textsf{C18}:\textsf{C22} \,\, 100:0 & Good homeotropic \\
\hline
\hline
\end{tabular}
\end{center}
\end{table}
%
%
%

From Figure \ref{speed} and Table \ref{alignment} we observe a connection
between the quality of the homeotropic alignment and the speed
of expansion of the homeotropic domains: for the higher speeds
we also have the best alignment.
Indeed, the higher speed implies a stronger anchoring which in turn gives a 
better alignment.
We also observe that in general mono-layers containing a high fraction
of \textsf{C18} give a better alignment.
The alignment given by \textsf{C22} is not very well 
defined,
but the addition of a small fraction of \textsf{C18} gives already a
homeotropic alignment, even thoug not very good.
Increasing the quantity of \textsf{C18} the homeotropic alignment improves
and reaches the best quality for the \textsf{C18}:\textsf{C22} ratios
of 50:50 (Figure \ref{conosco50}) and 60:40.
Also, the pure \textsf{C18} mono-layers gives a good 
homeotropic alignment.

\subsection*{Anchoring transition}
On heating, we observed a first-order
anchoring transition\cite{KomSteGabPug94,FazKomLag97a}
in a very narrow temperature
range, just below the clearing point.
At the transition a set of bright circular domains 
with dark crosses appear in the sample;
at constant temperature, they grow and coalesce, 
forming larger domains.
The appereance of these domains between crossed polarisers
is consistent with a degenerated tilted
orientation of the \textsf{NLC} molecules, or conical 
anchoring, also expected in the case of \textsf{LB} aligning 
films\cite{Jerome91}.

\begin{figure}
\begin{center}
\epsfig{file=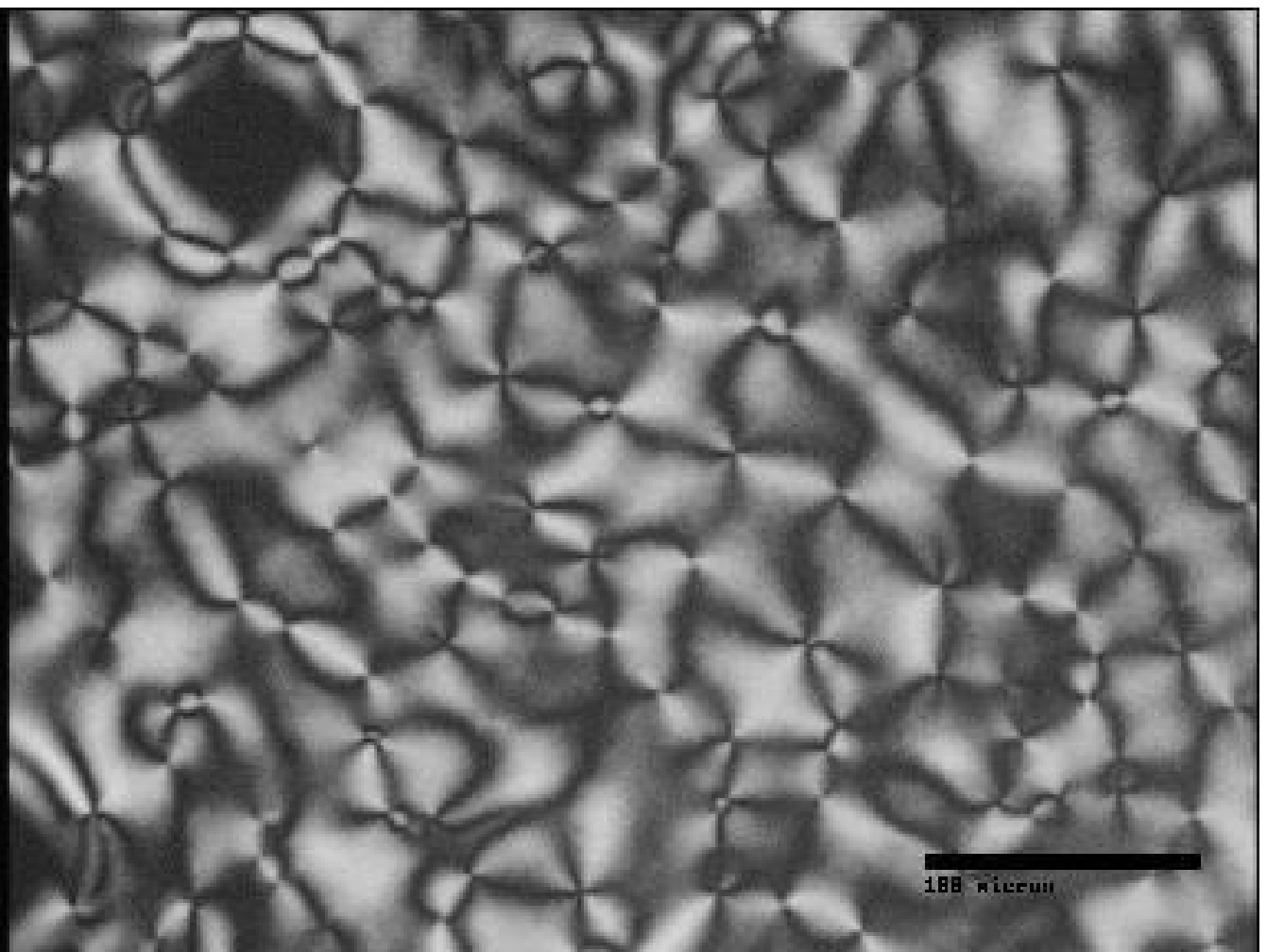,width=0.49\textwidth}
\epsfig{file=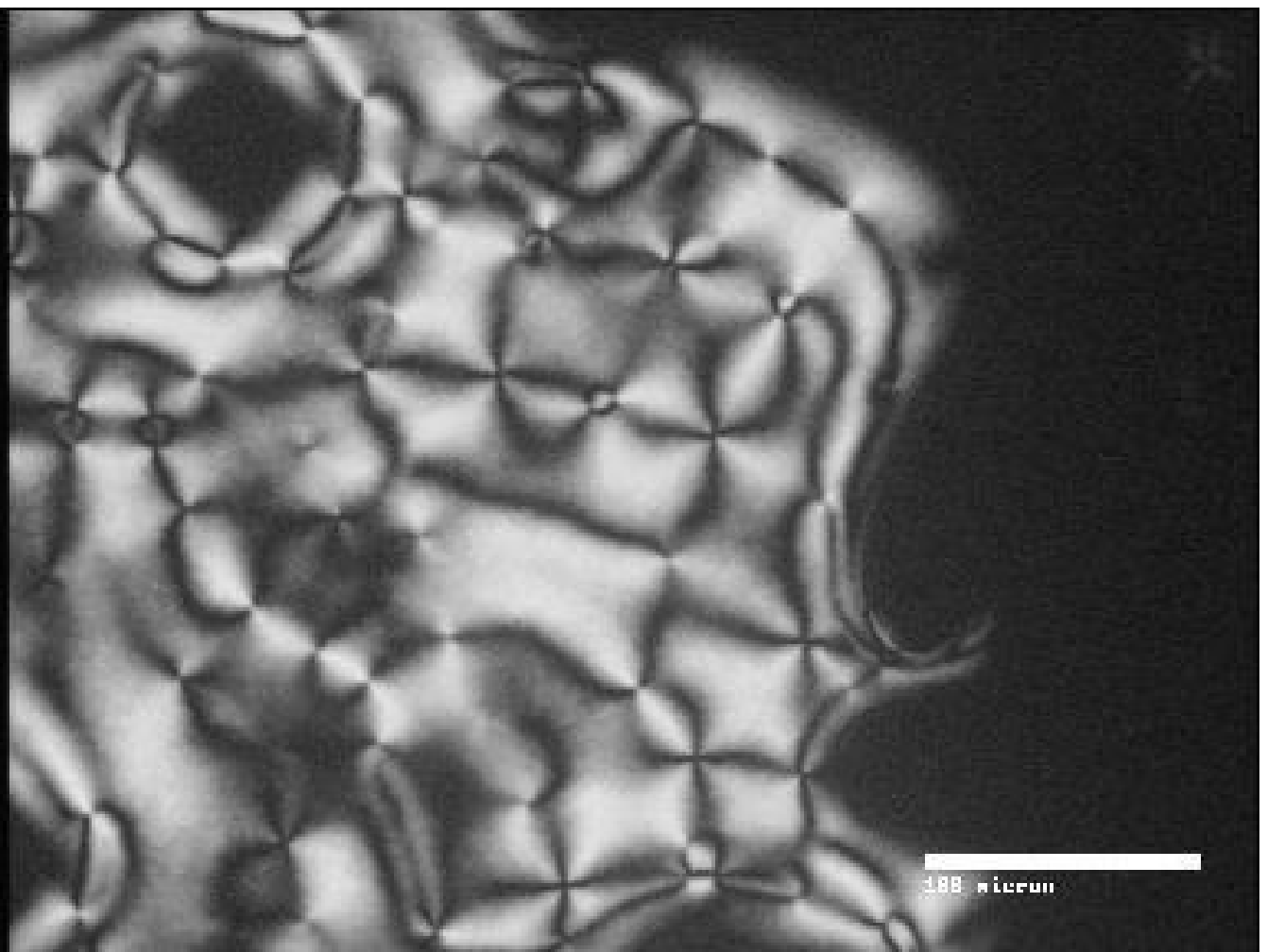,width=0.49\textwidth}
\caption{
\small{
Cell between crossed polarizers.
The aligning layer is \textsf{C18}/\textsf{C22} 20:80.
(a) On cooling from the isotropic phase, just below the clearing point,
we see a calssical schlieren texture
{\protect\cite{deGennes}}, as if we there were no aligning layer.
(b) 5$^{\circ}$C below the clearing point. 
The homeotropic alignment appears in domains (dark) which
extend to the whole cell.
\label{iso-nem}
}}
\end{center}
\end{figure}

On cooling from the isotropic phase the transition to the nematic 
homeotropic phase is different for the pure and the mixed mono-layers.
For the pure \textsf{C18} and \textsf{C22} mono-layers the bright domains
appear again and the transition to the homeotropic phase takes place inside
the domains\cite{KomSteGabPug94,FazKomLag97a}.
For all the mixed \textsf{C18}/\textsf{C22} mono-layers, on cooling from the
isotropic phase, a nematic, non-homeotropic phase appears just below the
clearing point (see Figure\ref{iso-nem}(a)).
It resembles the classical nematic phase in a cell without
aligning mono-layer:
it is characterised by domains were the director points in different
directions  (schlieren textures\cite{deGennes})
and it is stable in a temperature range of ca. 5$^{\circ}$C below
the clearing point.
On further cooling, domains with homeotropic alignment
appear in the sample and quickly
expand until the whole sample becomes homeotropic 
(see Figure \ref{iso-nem}(b)).

\begin{figure}
\begin{center}
\epsfig{file=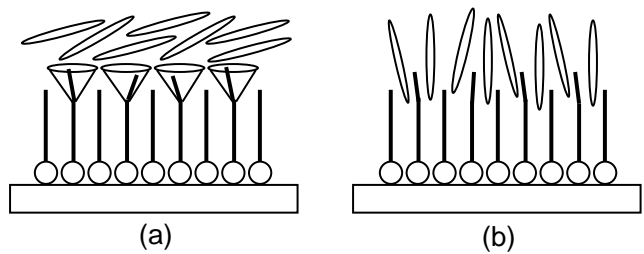,width=0.7\textwidth}
\caption{
\small{
Schematic view of the thermal motion of the upper segments of the aliphatic
chains{\protect\cite{ShaShi-incoll}}. 
(a) Just below the clearing point the thermal motion is large.
The cones represent the time-average space occupied by the thermal motion of
the segments.
In this situation there is not space for the liquid crystal molecules to
penetrate into the layer of the chains.
(b) More than 5$^{\circ}$C below the clearing point.
Now the space occupied by the chais in thermal motion is small enough to let
the liquid crystal molecules penetrate into the layer of the chains.
\label{coni}
}}
\end{center}
\end{figure}

A possible explanation for the different behaviour at the isotropic to nematic
phase transition is shown schematically in Figure 
\ref{coni}\cite{ShaShi-incoll}.
The portion of the behenic acid molecules above the height of the adjacent
stearic acid molecules will exhibit thermal motion which increases as the
temperature increases.
The phase transition from nematic to isotropic
is not much affected by this thermal
motion, being a transition from an ordered state to a  totally disordered one:
the thermal motion is in this case a help rather than an impediment.
On the contrary, at the phase transition from isotropic to nematic, 
this motion
prevents the recovering of the homeotropic alignment because 
the upper segments of the chains  
in motion occupy the space necessary for the liquid crystal molecules
to penetrate into the layer of the chains.
In this situation we have the nematic phase of the liquid crystal but
not the homeotropic alignment (Figure \ref{coni}(a)):
the schlieren texture will appear in the sample 
(see Figure \ref{iso-nem}(a))
because 
the \textsf{LB} film does not supply any preferred direction of alignment.
On cooling, the thermal fluctuations of the segments decrease  
and the liquid crystal molecules can penetrate into the layers
of the chains and restore the homeotropic alignment, as shown in 
Figure \ref{coni}(b).

\section*{Conclusions}
Alignment dynamics and properties of nematic liquid crystals on
pure and mixed stearic/behenic acids
Langmuir-Blodgett mono-layers have been investigated.
Mono-layers with a high \textsf{C18} fraction give a good homeotropic
alignment.
Increasing the fraction of \textsf{C22} the alignment deteriorates
and for pure \textsf{C22} there is no well defined alignment
at all in the cells.
The speed of the relaxation process from the splay-bend flow-induced
orientation to the homeotropic state, as well as the quality of
the homeotropic alignment, were found to depend on the composition of the
\textsf{LB} film.
They are also correlated, because a faster relaxation implies a 
stronger anchoring energy, which in turns gives a better alignment.
This has been also found in our experiment.

From the results we can conclude that the \textsf{LB} films
participate in the relaxation process in accordance with a model in which the
chains are mobile and can be deformed by the \textsf{LC} flow:
once the flow has ceased, the chains relax to their equilibrium state
with a speed which depends on the mono-layer composition.
In the equilibrium state the chains are in the upright position and move
only because of thermal fluctuations. 
At room temperature these fluctuations do not disturb the homeotropic
alignment, but in the case of mixed mono-layers 
on cooling from the isotropic phase they prevent the
transition from the isotropic directly to the nematic-homeotropic phase.
In fact, since the chains have different lengths, the upper
segments of the longest chains undergo strong fluctuations and do not
leave space for the liquid crystal molecules to penetrate
into the \textsf{LB} layer.

\bibliography{journals2,LCbibl}
\end{document}